\documentclass[3p,times]{elsarticle}

\usepackage{ecrc}

\volume{00}

\firstpage{1}

\journalname{Nuclear Physics A}

\runauth{K.F. Read for the PHENIX Collab.}

\jid{nupha}

\usepackage{amssymb}
\usepackage[figuresright]{rotating}

\newcommand{\pt}{{p_T}}

\newcommand{\raa}{{R_{\rm AA}}}

\newcommand{\sqs}{{\sqrt{s}}}
\newcommand{\snn}{{\sqrt{s_{\rm NN}}}}
\newcommand{\pp}{{$p$$+$$p$}}
\newcommand{\cucu}{{{\rm Cu}+{\rm Cu}}}

\newcommand{\ncol}{{\langle N_{\rm coll}\rangle}}

\newcommand{\nf}{{N_F}}
\newcommand{\nc}{{N_C}}
\renewcommand{\ni}{{N_I}}

\begin{document}

\begin{frontmatter}

%% Title, authors and addresses

%% use the tnoteref command within \title for footnotes;
%% use the tnotetext command for the associated footnote;
%% use the fnref command within \author or \address for footnotes;
%% use the fntext command for the associated footnote;
%% use the corref command within \author for corresponding author footnotes;
%% use the cortext command for the associated footnote;
%% use the ead command for the email address,
%% and the form \ead[url] for the home page:
%%
%% \title{Title\tnoteref{label1}}
%% \tnotetext[label1]{}
%% \author{Name\corref{cor1}\fnref{label2}}
%% \ead{email address}
%% \ead[url]{home page}
%% \fntext[label2]{}
%% \cortext[cor1]{}
%% \address{Address\fnref{label3}}
%% \fntext[label3]{}

\dochead{}
%% Use \dochead if there is an article header, e.g. \dochead{Short communication}

\title{Open Heavy Flavor Production at Forward Angles in PHENIX}

%% use optional labels to link authors explicitly to addresses:
%% \author[label1,label2]{<author name>}
%% \address[label1]{<address>}
%% \address[label2]{<address>}

\author{K.F. Read\fnref{doebyline} for the PHENIX Collaboration}

\address{Oak Ridge National Laboratory, Oak Ridge, TN 37831, USA}
\ead{readkf@ornl.gov}
\fntext[doebyline]{Research sponsored by the Office of Nuclear Physics, 
U.S. Department of Energy}

\begin{abstract}
%% Text of abstract

The measurement of the nuclear modification factor ($R_{\rm AA}$) for heavy-flavor
production in heavy-ion collisions tests predictions for cold- and
hot-nuclear-matter effects. Heavy-flavor production in \pp\ collisions
tests pQCD calculations and serves as a reference for understanding
heavy-flavor production in heavy-ion collisions. Using the PHENIX
muon-arm spectrometers, the transverse momentum spectra of inclusive muon candidates
are measured for \pp\ and $\cucu$ collisions at $\snn =
200$\,GeV. After subtracting backgrounds, we obtain the measured
invariant yields of negative muons from the decay of heavy flavor
mesons. For \pp\ collisions, we measure the charm-production cross
section integrated over $\pt$ and in the rapidity range $1.4<y<1.9$ to
be $d\sigma_{c\bar{c}}/dy = 0.139\pm 0.029\ {\rm
(stat)\,}^{+0.051}_{-0.058}\ {\rm (syst)}$~mb. This result is compared
to a recent FONLL calculation and to a PHENIX measurement at
mid-rapidity.  For $\cucu$ collisions, we measure the $R_{\rm AA}$ for
heavy-flavor muons in three centrality bins for $1<\pt<4$~GeV/$c$,
with suppression observed for central collisions. We compare our
measurement for central collisions to a recent theoretical prediction.

\end{abstract}

\begin{keyword}
%% keywords here, in the form: keyword \sep keyword
PHENIX \sep heavy ions \sep heavy flavor

%% MSC codes here, in the form: \MSC code \sep code
%% or \MSC[2008] code \sep code (2000 is the default)

\end{keyword}

\end{frontmatter}

%%
%% Start line numbering here if you want
%%
% \linenumbers

%% main text
\section{Introduction}
\label{secintro}

Heavy quarks are produced in the early stages of heavy-ion collisions
and, therefore, are a probe of the hot dense partonic matter produced.
Accurate measurements serve as a critical test and constraint for
predictions concerning the energy loss mechanism of partons moving
through such novel nuclear matter. This remains a priority for the
field of heavy-ion collision physics today. Measurement of heavy quark
production in $p$+$p$ collisions tests perturbative Quantum
Chromodynamics (pQCD) calculations, in addition to serving as a
reference for understanding production in heavy ion
collisions.

The nuclear modification factor $\raa$ for $\cucu$ collisions using a
$p$+$p$ reference is defined to be
\begin{equation}\label{eq:modRAA}
R_{\rm AA} = \frac{1}{\ncol} \frac{d^2N_{\cucu}/d\pt d\eta}
{d^2N_{pp}/d\pt d\eta},
\end{equation}

\noindent where $\ncol$ is the average number of
nucleon-nucleon collisions in the $\cucu$ collision based on a
simple description of the nucleus~\cite{Miller:2007ri},
averaged over a given centrality bin. Collisions are classified
experimentally according to
centrality~\cite{Miller:2007ri}, the percentage of the 
total nuclear inelastic cross section, with 0\% representing the most
head-on collisions.

In the absence
of nuclear effects, $\raa$ is expected to be equal to unity. Values
less than unity are referred to as ``suppression.''
Heavy quark production
is expected to be modified by cold-nuclear matter effects, such as
shadowing and initial-state energy loss, which can effect $\raa$.
It is important to complement existing measurements for open heavy
flavor production at mid-rapidity with measurements at forward
rapidities to fully test theoretical models.

\section{Methodology}
\label{secmethod}

\begin{figure}
\centering
\includegraphics[width=0.45\linewidth]{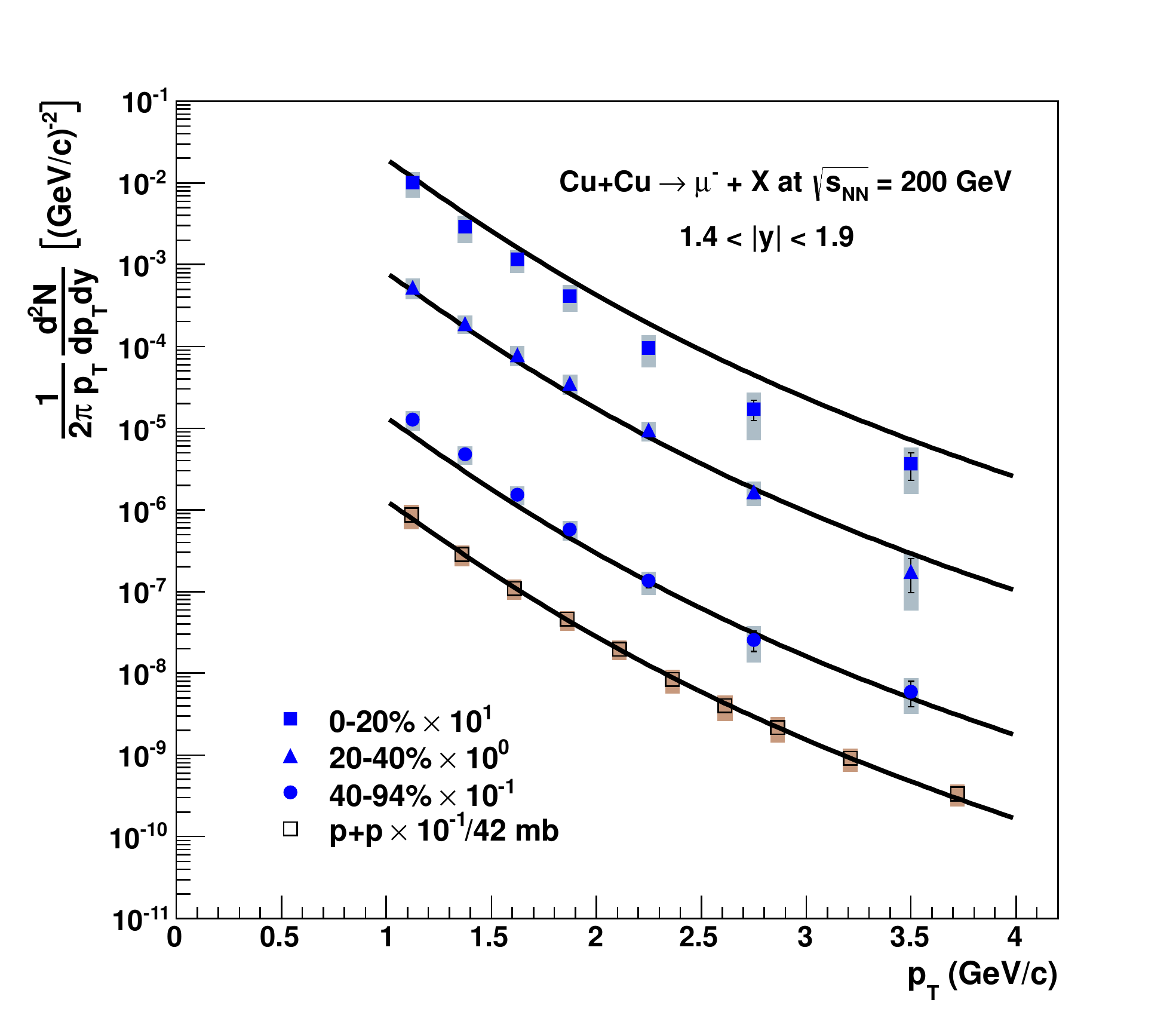}
\includegraphics[width=0.45\linewidth]{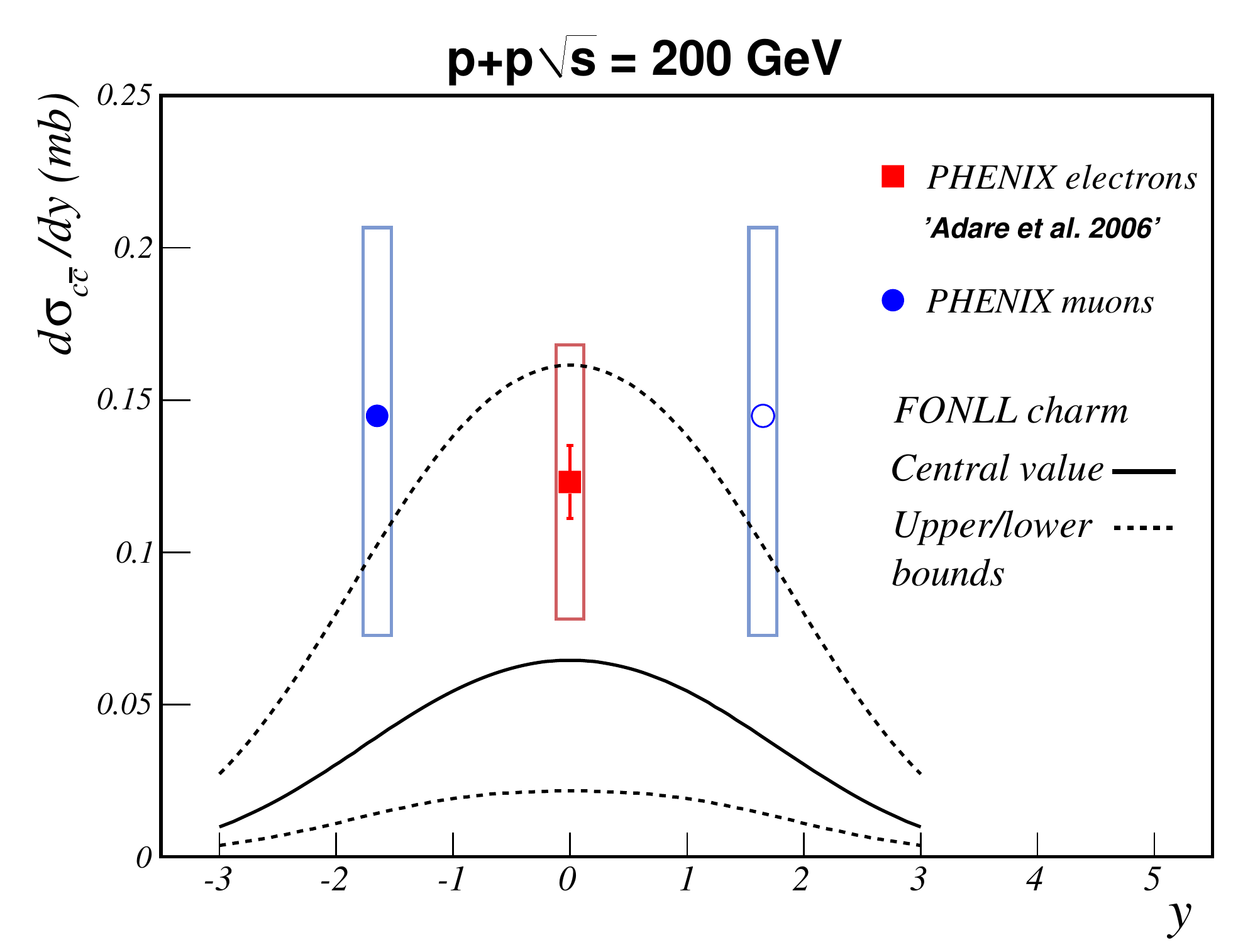}
\caption{
Left: Invariant yields of negative muons produced from heavy-flavor meson
decay as a function of $\pt$ in $p$$+$$p$ collisions at $\sqs = 200$
GeV (open squares) and in $\cucu$ collisions at $\snn = 200$\,GeV for
three different centrality bins (40--94\%, 20--40\% and 0--20\%),
scaled by powers of ten for clarity (filled symbols)~\cite{Adare:2012px}.
Solid lines are a fit to the \pp\ yield described in the text~\cite{Kaplan:1978},
scaled by the corresponding average number of binary collisions $\ncol$.
Right: $c\overline{c}$ production cross section as a function of
rapidity in $p$$+$$p$ collisions at $\sqs = 200$\,GeV, measured using
semileptonic decay muons at forward rapidity
(circles)~\cite{Adare:2012px} and electrons at
mid-rapidity (square)~\cite{Adare:2006hc}. Curves represent
a theoretical prediction~\cite{fonll:1998,fonll:2001}.
}
\label{fig:yields}
\end{figure}

This analysis is based on measurements by the PHENIX Experiment of
\pp\ and $\cucu$ collisions at a center-of-mass energy per
nucleon-nucleon collision of $\snn = 200$\,GeV. The south and north
PHENIX muon arm spectrometers cover the pseudorapidity range
$-2.2<\eta<-1.2$ and $1.2<\eta<2.4$, respectively, over the full
azimuth. Each muon arm contains a Muon Identifier (MuID) consisting of
interleaved planes of steel absorber and Iarocci tubes, with the five
instrumented ``gaps'' numbered from 0 to 4. Beam-Beam Counters (BBC), composed
of arrays of quartz \v{C}erenkov detectors on either side of the 
interaction point, provide a minimum bias trigger.

The double differential invariant yield for muons from the semileptonic
decay of heavy flavor mesons is:

\begin{equation}\label{eq:smyield}
\frac{d^{2}N^{\mu}}{2 \pi \pt d\pt d\eta} 
= \frac{1}{2\pi\pt\Delta\pt \Delta\eta}\frac{\ni-\nc-\nf}
{N_{\rm evt}\epsilon_{\rm BBC}^{c\overline{c}\rightarrow \mu}A\epsilon}
\end{equation}

\noindent where $\ni$ is the total number of muon candidates in the
bin that reach the last gap of the MuID (Gap 4) and pass all track
selection criteria; $\nc$ is the number of tracks corresponding to the
irreducible hadronic background estimated using a hadron cocktail;
$\nf$ is the estimated number of surviving misreconstructed tracks;
$N_{\rm evt}$ is the number of events; $A\epsilon$ is the detector
acceptance and efficiency correction; and $\epsilon_{\rm
BBC}^{c\overline{c}\rightarrow \mu}$ is the BBC trigger efficiency for
events with a heavy-flavor muon at forward rapidity.
The signal-to-background ratio is better for negative muons than positive
muons because antiprotons and negative kaons are more suppressed
than their positive counterparts by the Muon Tracker front absorbers. Therefore,
only negative muon candidates (and background estimates) are considered in 
this analysis~\cite{Adare:2012px}.

A data-driven Monte Carlo ``cocktail'' is used to predict/simulate the
irreducible hadronic background $\nc$. The cocktail consists of a
collection of individual simulated primary hadrons which are weighted
to match RHIC data. Specifically, the transverse momentum spectrum for
neutral pions is adjusted to agree with PHENIX central arm
measurements extrapolated to our rapidity region. Other hadrons are
thrown so that they conform to hadron-to-pion ratios measured by
experiments at RHIC. In order to account for any degradation of
reconstructed track quality in the $\cucu$ analysis, we embed the
hadron cocktail tracks in real $\cucu$ events.
The ``cocktail'' is propagated via GEANT through a complete simulation
of the PHENIX muon arms. In order to explore sensitivity to the
simulated propagation through the thick material of the MuID, we
produced background estimates using both the {\sc FLUKA} and {\sc GHEISHA} hadron
shower codes, with a range of modified hadron-iron interaction cross
sections~\cite{Adare:2012px}.
Tracks which stop at MuID Gap 2 or 3 with no hit further downstream
(Gap 3 or 4, respectively) have a characteristic distribution of {\it
longitudinal} momentum. Decay muons exhibit a sharp peak due to
electromagnetic energy loss, but hadrons exhibit a broad shoulder. A
highly purified sample of {\it hadrons} can be obtained from such
tracks by requiring $p_z > 3$ GeV/c. The reliance on initial
distributions for the hadron cocktail is significantly minimized by
subsequent ``tuning'' which further adjusts the hadron weights to
match the measurements for {\it hadrons} from MuID Gap 2 and Gap 3.

Simulation studies show that the number of decay muons that reach MuID
Gap 4 linearly increases with the longitudinal distance of the vertex
from the respective muon arm. Hadrons do not exhibit this dependence.
The inclusive sample of tracks reaching MuID Gap 4 will exhibit a
linear dependence of $dN_\mu/dz_{\rm BBC}$ on $z_{\rm BBC}$, the event vertex
longitudinal position measured by the BBC. The
weights of the hadron cocktail can be adjusted to match this slope for
each arm for different transverse momentum bins.

The number of tracks in the hadron cocktail that satisfy all
selection criteria is $\nc$, which we subtract from the inclusive yield
$\ni$ as indicated in Equation~\ref{eq:smyield}. After correcting for
acceptance and efficiencies, we obtain the measured invariant yield of
negatively charged muons from the decay of heavy-flavor mesons in \pp\
collisions at $\sqs = 200$\,GeV and for three centrality bins of
$\cucu$ collisions at $\snn = 200$ GeV.  In order to obtain the
nuclear modification factor and fully exploit the cancellation of
common systematic errors, we use Equation~\ref{eq:modRAA} for each bin
of transverse momentum and for a common hadron cocktail
optimization. This is possible in an experiment which measures its own
reference data for computation of a nuclear modification factor.

\section{Results}
\label{secresults}

\begin{figure}
\centering
\includegraphics[width=0.4\linewidth]{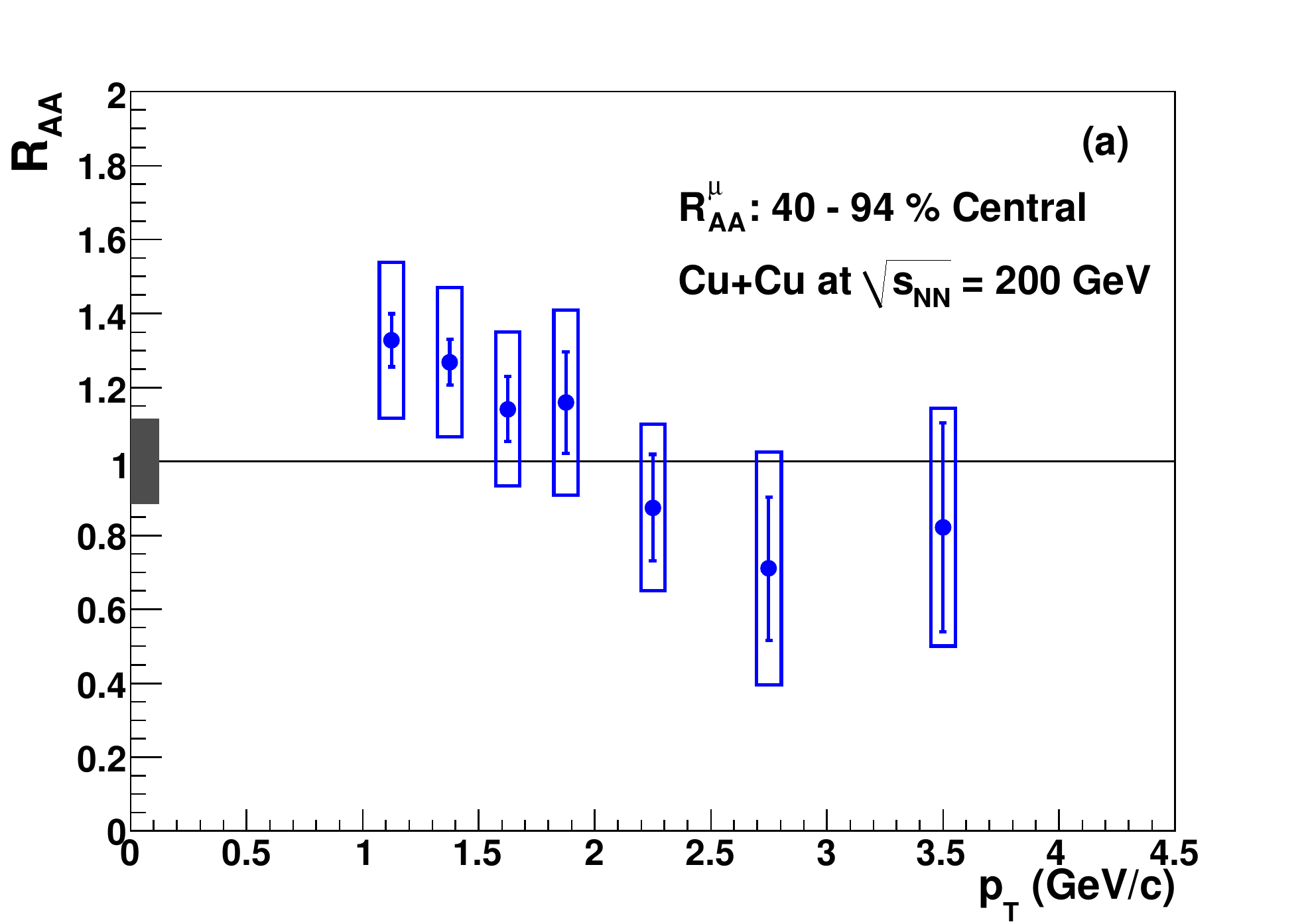} 
\includegraphics[width=0.4\linewidth]{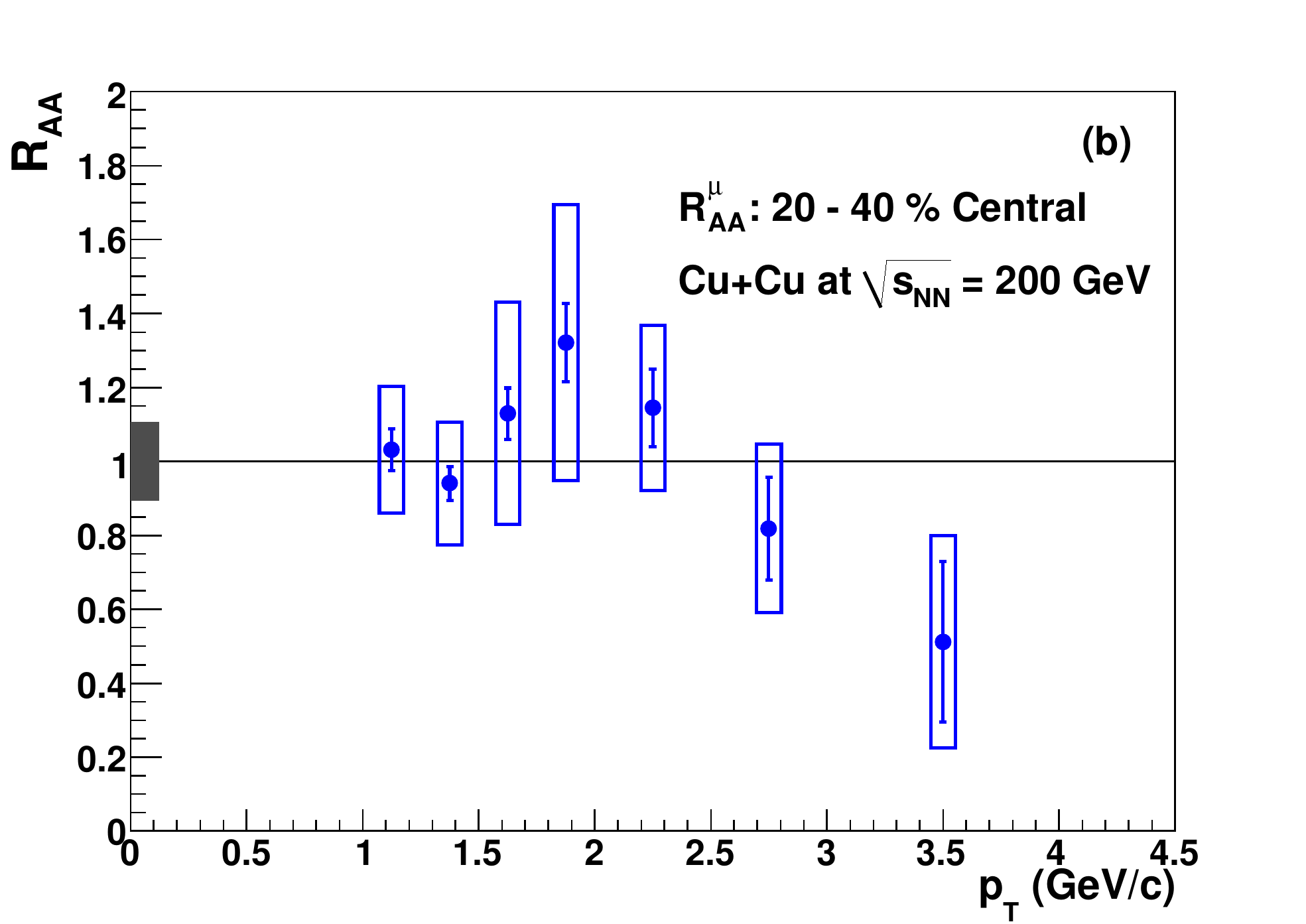} 
\includegraphics[width=0.4\linewidth]{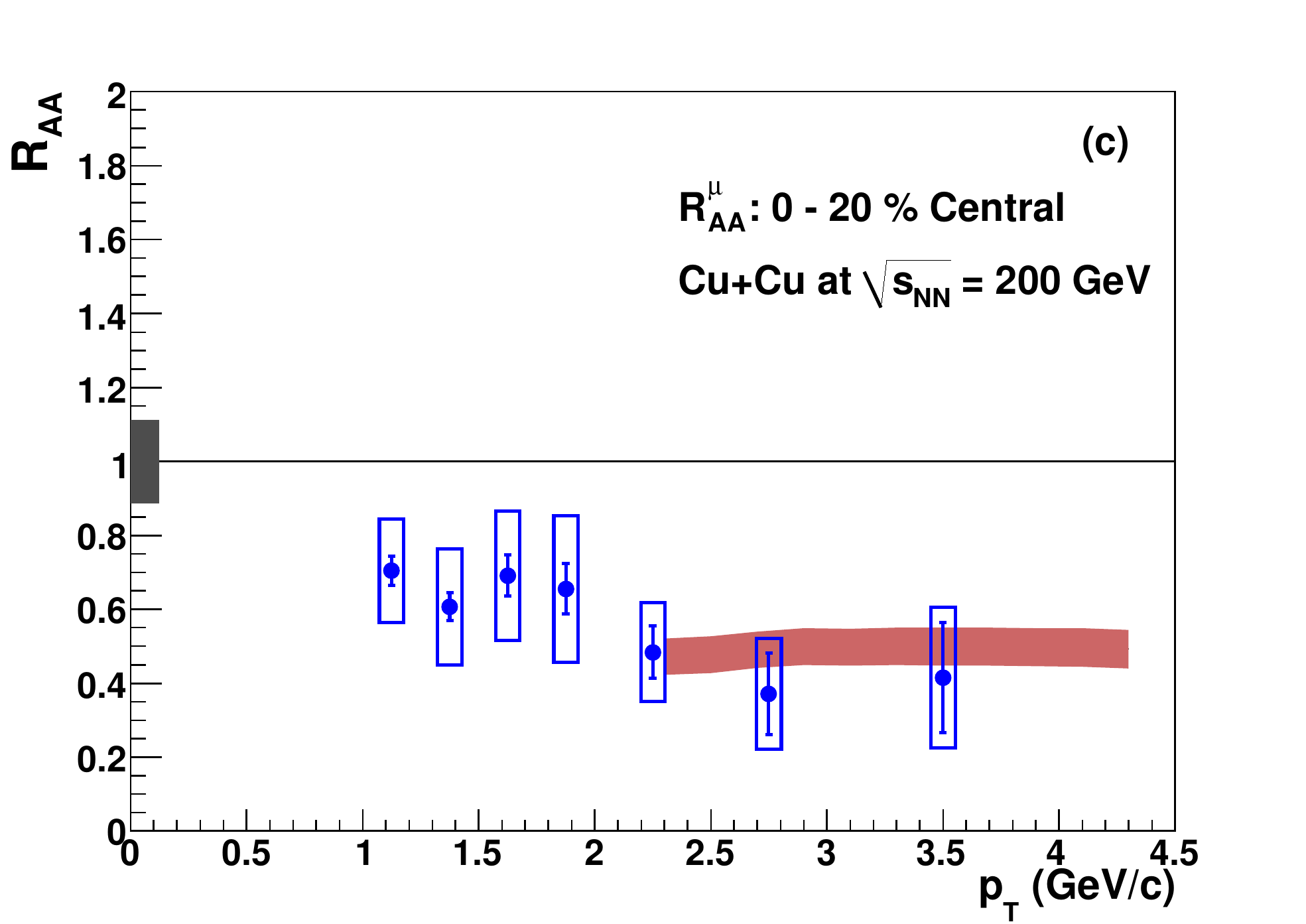}
\caption{
(Color online) Transverse momentum distribution of $\raa$ for negative muons from
heavy-flavor meson decay in $\cucu$ collisions at $\snn=200$\,GeV for the centrality
bins: 40--94\% (a), 20--40\% (b), and 0--20\% (c)~\cite{Adare:2012px}.
The band in (c) is a theoretical prediction
from~\cite{Vitev:private,Sharma:2009hn}.}
\label{fig:raafinal}
\end{figure}

The measured invariant yields of negatively charged muons from the
decay of heavy-flavor mesons in \pp\ collisions at $\sqs = 200$\,GeV
and for three centrality bins of $\cucu$ collisions at $\snn = 200$
GeV are shown in the left panel of Figure~\ref{fig:yields} scaled by powers of 10
for clarity.  The solid lines are a fit to the \pp\ yield
using the function \mbox{$A[1+(\pt/B)^{2}]^{-4.2}$} based on~\cite{Kaplan:1978},
scaled by the corresponding average number of binary collisions $\ncol$
for the $\cucu$ measurements. The contribution from
bottom meson decay is estimated to be negligible for this transverse
momentum region.
In order to estimate the full charm cross section for \pp\ collisions,
we express the \pp\ invariant yield as a differential cross section
$d\sigma_{\mu^-}/dy$ after extrapolating the measurement to $p_{\rm T}
= 0$ GeV/c using fixed-order-plus-next-to-leading-log
(FONLL)~\cite{fonll:1998,fonll:2001} calculations. The charm-production
cross section $d\sigma_{c\bar{c}}/dy$ is then computed using:
\begin{equation}
d\sigma_{c\bar{c}}/dy = 
\frac{1}{BR(c \rightarrow \mu)} \cdot \frac{1}{C_{l/D}} 
\cdot \frac{d\sigma_{\mu^-}}{dy}
\end{equation}
where $BR(c \rightarrow \mu)$ is the total charm to muon branching
ratio and $C_{l/D}$ is a kinematic correction factor based on the
FONLL calculation. We obtain a charm-production cross section
integrated over $\pt$ and in the rapidity range $1.4<y<1.9$ of
$d\sigma_{c\bar{c}}/dy = 0.139\pm 0.029\ {\rm
(stat)\,}^{+0.051}_{-0.058}\ {\rm
(syst)}$~mb. The right panel of Figure~\ref{fig:yields} shows the $c\overline{c}$
production cross section as a function of rapidity using this
measurement of muons at forward rapidity~\cite{Adare:2012px}
compared to a PHENIX measurement using electrons at
mid-rapidity~\cite{Adare:2006hc}.

Figure~\ref{fig:raafinal} shows the transverse momentum distribution
of $\raa$ for negative muons from heavy-flavor meson decay in $\cucu$
collisions for three centrality bins~\cite{Adare:2012px}. The (red)
band in Figure~\ref{fig:raafinal} shows a recent theoretical
prediction~\cite{Vitev:private,Sharma:2009hn} for $y=1.65$ and $\pt >
2.5$\,GeV/c which includes the effects of heavy quark energy loss,
in-medium heavy meson dissociation, and cold nuclear matter effects
such as shadowing and initial state energy loss of the incoming
partons.

\section{Discussion and Conclusions}
\label{secconclusions}

Negative muons from the decay of heavy flavor mesons in \pp\ collisions
at $\sqs = 200$\,GeV are used to measure the charm-production cross
section integrated over $\pt$ and in the rapidity range
$1.4<y<1.9$. The result is $d\sigma_{c\bar{c}}/dy = 0.139\pm 0.029\
{\rm (stat)\,}^{+0.051}_{-0.058}\ {\rm (syst)}$~mb. This result is
compatible with a FONLL calculation within experimental and
theoretical scale uncertainties. Negative muons from heavy-flavor meson
decay have also been measured in $\cucu$ collisons at $\snn=200$\,GeV
for the same rapidity and momentum range, and for 3 bins of
centrality.  Reasonable agreement is observed between the invariant
yields for the peripheral ($40 - 90$\%) and mid-central ($20 - 40$\%)
collisions and scaled fits to the invariant yield for \pp\ collisions;
however, a suppression is observed at higher $\pt$ for the central
($0 - 20$\%) $\cucu$ collisions. This effect is quantified by the
reported nuclear modification factors $R_{\rm AA}$ measured in
three centrality bins. We observe significant suppression of
heavy-flavor muon production in central $\cucu$ collisions ($0 -
20$\%), with the largest effect for $\pt > 2$\,GeV/c.
The measurement for central collisions is consistent with a
recent theoretical prediction for this rapidity and transverse
momentum range~\cite{Vitev:private,Sharma:2009hn}.

The PHENIX Experiment has installed new inner silicon vertex detectors
which will allow separation of charm and bottom contributions in future
measurements. Such measurements will constrain predictions and help
guide theoretical understanding as part of an improving understanding
of QCD matter at extreme energy densities.

%% The Appendices part is started with the command \appendix;
%% appendix sections are then done as normal sections
%% \appendix

%% \section{}
%% \label{}

%% References
%%
%% Following citation commands can be used in the body text:
%% Usage of \cite is as follows:
%%   \cite{key}         ==>>  [#]
%%   \cite[chap. 2]{key} ==>> [#, chap. 2]
%%

%% References with BibTeX database:

\bibliographystyle{elsarticle-num}
\bibliography{hp2012-KennethREAD}

%% Authors are advised to use a BibTeX database file for their reference list.
%% The provided style file elsarticle-num.bst formats references in the required Procedia style

%% For references without a BibTeX database:

% \begin{thebibliography}{00}

%% \bibitem must have the following form:
%%   \bibitem{key}...
%%

% \bibitem{}

% \end{thebibliography}

\end{document}